\documentclass[journal]{IEEEtran}
\usepackage{fancyhdr}
\usepackage{amsmath,bm}
\usepackage{amssymb}
\usepackage{graphicx}
\usepackage{booktabs}
\usepackage{amsfonts}
\usepackage{multirow}
\usepackage{algorithm}
\usepackage{algorithmic}
\usepackage{flushend}
\usepackage{mathrsfs}
\usepackage{color}

\usepackage{exscale}
\usepackage{relsize}

\usepackage{subfigure}
\usepackage{cite}

\newtheorem{theorem}{\textbf{Theorem}}

\ifCLASSINFOpdf
\else
\fi

\begin{document}
%
\title{Spatial Modulation Aided Layered Division Multiplexing: A Spectral Efficiency Perspective}
%


\author{Yue~Sun,~\IEEEmembership{Student~Member,~IEEE},
        Jintao~Wang,~\IEEEmembership{Senior~Member,~IEEE},
        Changyong~Pan,~\IEEEmembership{Senior~Member,~IEEE},
        Longzhuang~He,~\IEEEmembership{Student~Member,~IEEE},
        and Bo~Ai,~\IEEEmembership{Senior~Member,~IEEE}
\thanks{
Y. Sun, J. Wang, C. Pan and L. He are with the Department of Electronic Engineering and Tsinghua National Laboratory for Information Science and Technology (TNList), Tsinghua University, Beijing, 100084, China (email: suny15@tsinghua.edu.cn).

B. Ai is with the State Key Laboratory of Rail Traffic Control and Safety, Beijing Jiaotong University, Beijing, 100044, China (e-mail: aibo@ieee.org).

This work was supported in part by the National Key R\&D Program of China (Grant No. 2016YFB1200102-04) and the National Natural Science Foundation of China (Grant No. 61471221 and No. 61471219).
}}

\maketitle

\begin{abstract}
In this paper, spatial modulation (SM) is introduced to layered division multiplexing (LDM) systems for enlarging the spectral efficiency over broadcasting transmission. Firstly, the SM aided LDM (SM-LDM) system is proposed, in which different layered services utilize SM for terrestrial broadcasting transmission with different power levels. Then a spectral efficiency (SE) analysis framework for SM-LDM systems is proposed, which is suitable for the SM-LDM systems with linear combining. Moreover, the closed-form SE lower bound of SM-LDM systems with maximum ratio combining (MRC) is derived, which is based on this framework. Since the theoretical SE analysis of single transmit antenna (TA) LDM systems with MRC and spatial multiplexing (SMX) aided LDM systems with MRC lacks a closed-form expression, the closed-form SE is also derived for these systems. Monte Carlo simulations are provided to verify the tightness of our proposed SE lower bound. Furthermore, it can be shown via simulations that our proposed SM-LDM systems always have a better SE performance than single-TA LDM systems, which can even outperform the SE of SMX aided LDM (SMX-LDM) systems.
\end{abstract}


\IEEEpeerreviewmaketitle

\section{Introduction}

\IEEEPARstart{L}{ayered} division multiplexing (LDM) technology is recently proposed to satisfy the rapidly increasing spectral efficiency (SE) demand of digital terrestrial television (DTT) transmission, which has been accepted in the Advanced Television Systems Committee (ATSC) 3.0 standard \cite{ATSC, ATSC_MIMO, LDM_ATSC, LDM_TAP, LDM_MIMO_C, LDM_BMSB}. As a non-orthogonal multiplexing technology, LDM simultaneously transmits different layered services at different power levels. Comparing with traditional time division multiplexing (TDM) and frequency division multiplexing (FDM), LDM has a higher SE, which is benefited from power allocation of different services \cite{LDM_MIMO_C}. Since different layers share the main part of physical layer modules, the LDM system only has a slightly higher complexity than the FDM or TDM system \cite{LDM_ATSC}.

For LDM systems, in most instances there are two layers, i.e., the upper layer (UL) and the lower layer (LL), and the UL is allocated with a higher power level than the LL \cite{LDM_TAP}. The UL delivers low data rate service for mobile receivers, and the FL delivers high data rate service for fixed receivers. Therefore, the UL and the LL are also referred to as mobile layer (ML) and fixed layer (FL), respectively. When detecting the ML service, the FL service is treated as additional interference, and when detecting the FL service, the ML service need to be firstly cancelled \cite{LDM_BMSB}.

Spatial modulation (SM) is proposed as a novel architecture of multiple-input multiple-output (MIMO) systems, which only activates one transmit antenna (TA) for delivering the constellation symbol in each time slot with only one radio frequency (RF) chain \cite{SM} \cite{SM_mgz}. Therefore, the information can be transmitted from both the spatial domain and constellation domain, and SM systems can achieve a better energy efficiency (EE) than traditional MIMO systems. In addition, with only one TA active in each time slot, SM has a more relaxed inter-antenna-synchronization (IAS) than traditional MIMO systems, and SM has no inter-channel interference (ICI) \cite{SM_mgz}.

SM systems can also be combined with other schemes, such as massive SM MIMO systems \cite{Massive_SM_ITA} \cite{Massive_SM_TVT}, non-orthogonal multiple access aided SM (SM-NOMA) systems \cite{NOMA_SM_WXS} and generalized spatial modulation (GenSM) aided millimeter wave (mmWave) systems \cite{Mmwave_SM_HLZ} \cite{Mmwave_SM_HLZ_TCOM}. In broadcasting transmission scenarios, SM systems are also introduced to obtain a better trade-off of SE and EE \cite{SM_TBC}\cite{SM_Jiao}. More specifically, in \cite{SM_TBC}, SM is combined with massive MIMO and orthogonal frequency division multiplexing (OFDM) in high speed train systems, and in \cite{SM_Jiao}, a block-sparse compressive sensing (BS-CS) based method is proposed for detection of GenSM with NOMA in terrestrial return channel.

However, to the best of our knowledge, there are no research about the SM system combined with LDM system. Therefore, in this paper, we combine the SM system with a two-layer LDM system, which is denoted as the SM aided LDM (SM-LDM) system. In this SM-LDM system, both the ML service and FL service utilize SM for terrestrial broadcasting transmission. The SE analysis framework of SM-LDM systems with linear combining is also proposed, in which the signal-to-interference-plus-noise-ratio (SINR) determined by specific combining schemes is the only variable of mutual information (MI). Moreover, the closed-form SE lower bound of SM-LDM systems with maximum ratio combining (MRC) is derived by calculating out the SINR value. In addition, since the derived SE of single-TA LDM systems with MRC and spatial multiplexing (SMX) aided LDM systems with MRC lack the closed-form expressions \cite{LDM_TAP}\cite{LDM_MIMO_C}, we also derive the closed-form SE of these systems.

The organization of this paper is summarized as follows. In Section \uppercase\expandafter{\romannumeral2}, the system model of our proposed SM-LDM is introduced. In Section \uppercase\expandafter{\romannumeral3}, the SE analysis framework of SM-LDM systems with linear combining is proposed. In Section \uppercase\expandafter{\romannumeral4}, the closed-form SE lower bound of SM-LDM systems with MRC is derived by calculating out the SINR. Section \uppercase\expandafter{\romannumeral5} presents the Monte Carlo simulation results to show the tightness of our proposed SE lower bound of SM-LDM systems with MRC, and the comparison between SM-LDM systems and other LDM schemes are also provided in this section. Finally, Section \uppercase\expandafter{\romannumeral6} concludes this paper.

\emph{Notations}: In this paper, the uppercase and lowercase boldface letters represent matrices and column vectors, respectively. The operators $| \cdot |$, $(\cdot)^T$, $(\cdot)^H$ and $\|(\cdot)\|$ indicate the absolute function, transposition, conjugate transposition and Frobenius norm, respectively. The abbreviations $\text{det}(\mathbf{A})$ and $\mathbf{A}(i,j)$ denote the determinant of matrix $\mathbf{A}$ and the component of $\mathbf{A}$ in $i$-th row and $j$-th column, respectively. The abbreviation diag$(\mathbf{x})$ represents a diagonal matrix with diagonal elements $\mathbf{x}$. $\mathcal{P}(\cdot)$ denotes the probability density function, $\mathcal{CN}(\boldsymbol{\mu}, \mathbf{\Sigma})$ denotes a circularly symmetric multi-variate complex Gaussian distribution with mean $\boldsymbol{\mu}$ and covariance $\mathbf{\Sigma}$, and $\mathcal{CN}(\mathbf{x}; \boldsymbol{\mu}, \mathbf{\Sigma})$ denotes the probability density function (PDF) of the random vector $\mathbf{x} \sim \mathcal{CN}(\boldsymbol{\mu}, \mathbf{\Sigma})$.

\section{System Model}

\begin{figure}
  \centering
  \includegraphics[width=0.45\textwidth]{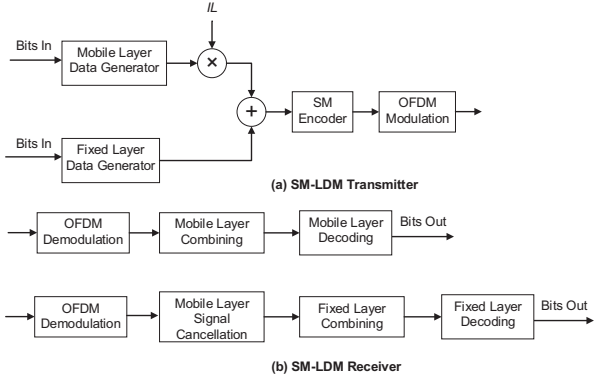}
  \caption{Transmitter and receiver for the two-layer SM-LDM system.}\label{system_SM_LDM}
\end{figure}

In this paper, a two-layer SM-LDM downlink model for terrestrial broadcasting transmission is introduced, which can also be easily extended to a multi-layer SM-LDM downlink model. As shown in Fig. \ref{system_SM_LDM} (a), at the transmitter, firstly the constellation symbols of ML and FL are separately generated in frequency domain, and the active antennas of two layers are also determined. Then the injection level, \textit{IL}, is introduced to control the power allocation between the two layers \cite{LDM_MIMO_C}. After that, for SM-OFDM scheme, each subcarrier relies on one TA \cite{SM_TBC}, so each constellation symbol in frequency domain is allocated with one active antenna. Therefore, in our proposed two-layer SM-LDM system, the transmitted symbol in frequency domain can be denoted as follows:
\begin{equation}
\mathbf{x} = \sqrt{\rho_\text{ml}}\mathbf{x}_\text{ml} + \sqrt{\rho_\text{fl}}\mathbf{x}_\text{fl},
\label{T_symbol}
\end{equation}
where $\rho_\text{ml}$ and $\rho_\text{fl}$ denote the transmit power of ML and FL, respectively. Besides, $\mathbf{x}_\text{ml} \in \mathbb{C}^{N_\text{t} \times 1}$ and $\mathbf{x}_\text{fl} \in \mathbb{C}^{N_\text{t} \times 1}$ denote the frequency-domain transmit symbol of ML and FL, respectively. The number of TAs at the transmitter is denoted as $N_\text{t}$. Since the ML is allocated with a higher power than FL, we have:
\begin{equation}
\rho_\text{ml}+\rho_\text{fl} = P_\text{u},\ \  \rho_\text{ml}/\rho_\text{fl} = \textit{IL}, \ \ \textit{IL} > 0\ \text{dB},
\end{equation}
where $P_\text{u}$ denotes the total transmit power. Aided by the property of SM, the $\mathbf{x}_\text{ml}$ and $\mathbf{x}_\text{fl}$ can be denoted as follows:
\begin{equation}
\mathbf{x}_\text{ml} = s_\text{ml}\mathbf{a}_\text{ml}, \ \ \mathbf{x}_\text{fl} = s_\text{fl}\mathbf{a}_\text{fl},
\end{equation}
where $s_\text{ml}$ and $s_\text{fl}$ denote the constellation symbols of ML and FL, respectively. $\mathbf{a}_\text{ml} = [0,\ldots,0,1,0,\ldots,0]^{T} \in \mathbb{C}^{N_\text{t} \times 1}$ and $\mathbf{a}_\text{fl} = [0,\ldots,0,1,0,\ldots,0]^{T} \in \mathbb{C}^{N_\text{t} \times 1}$ denote the active antenna of ML and FL, respectively. For both $\mathbf{a}_\text{ml}$ and $\mathbf{a}_\text{fl}$, only one element representing the active antenna is equal to $1$, and other elements are equal to $0$.

At the mobile receiver, as shown in Fig. \ref{system_SM_LDM} (b), the FL symbol is regarded as additional interference, and we denote $N_\text{rm}$ as the number of receive antennas (RAs) in ML. Thus the received symbol can be denoted as follows:
\begin{equation}
\mathbf{y}_\text{ml} = \mathbf{H}_\text{ml}\mathbf{x} + \mathbf{n}_\text{ml} = \mathbf{H}_\text{ml}\left(\sqrt{\rho_\text{ml}}\mathbf{x}_\text{ml} + \sqrt{\rho_\text{fl}}\mathbf{x}_\text{fl}\right) + \mathbf{n}_\text{ml},
\label{ML_matrix}
\end{equation}
where $\mathbf{H}_\text{ml} \in \mathbb{C}^{N_\text{rm}\times N_\text{t}}$ represents the frequency-domain channel matrix between the transmitter and the ML receiver. Assuming a Wide Sense Stationary (WSS) Rayleigh fading channel \cite{LDM_TAP}, each element of $\mathbf{H}_\text{ml}$ is an independent and identically distributed (i.i.d.) Gaussian random variable with mean $0$ and variance $1$. In addition, $\mathbf{n}_\text{ml}\in\mathbb{C}^{N_\text{rm}\times 1}$
denotes the additive white Gaussian noise (AWGN) of ML with $\mathbf{n}_\text{ml} \sim \mathcal{CN}(\mathbf{0},\sigma_\text{ml}^2\mathbf{I})$, and $\sigma_\text{ml}^2$ is the noise variance of ML.

At the fixed receiver, $N_\text{rf}$ denotes the number of RAs, and the received symbol can be denoted as follows:
\begin{equation}
\mathbf{y}_\text{fl} = \mathbf{H}_\text{fl}\mathbf{x} + \mathbf{n}_\text{fl} = \mathbf{H}_\text{fl}\left(\sqrt{\rho_\text{ml}}\mathbf{x}_\text{ml} + \sqrt{\rho_\text{fl}}\mathbf{x}_\text{fl}\right) + \mathbf{n}_\text{fl},
\end{equation}
where $\mathbf{H}_\text{fl} \in \mathbb{C}^{N_\text{rf}\times N_\text{t}}$ is the channel matrix between the transmitter and the FL receiver in frequency domain, which can also be assumed as a WSS Rayleigh fading channel, so each element of $\mathbf{H}_\text{fl}$ is i.i.d. Gaussian random variable with $\mathbf{H}_\text{fl}(i,j)\sim \mathcal{CN}(0,1)$. $\mathbf{n}_\text{fl}\in\mathbb{C}^{N_\text{rf}\times 1}$ is the AWGN of FL with $\mathbf{n}_\text{fl}\sim\mathcal{CN}(\mathbf{0},\sigma_\text{fl}^2\mathbf{I})$, and $\sigma_\text{fl}^2$ is the noise variance of FL. The ML noise always has a higher power level than the FL noise, and thus we have $\sigma_\text{ml}^2 > \sigma_\text{fl}^2$.

When detecting the symbols of FL, as shown in Fig. \ref{system_SM_LDM} (b), before detecting the FL signal, the ML signal is firstly cancelled. The perfect ML signal cancellation is assumed, which is because in the typical scenarios of ATSC 3.0 system with LDM, the FL always has a much higher signal-to-noise-ratio (SNR) than that of ML \cite{LDM_TAP}. However, after the perfect ML signal cancellation, the cross-layer interference (CLI) still might be introduced because of the non-ideal channel estimation (CE). Fortunately, since a properly designed CE module can provide a CE mean square error (MSE) lower than $-30$ dB \cite{LDM_BMSB}, the CLI is not explicitly considered \cite{LDM_MIMO_C}. Therefore, after the ML signal cancellation, the received symbol of FL can be denoted as follows:
\begin{equation}
\mathbf{y}_\text{fl} = \sqrt{\rho_\text{fl}}\mathbf{H}_\text{fl}\mathbf{x}_\text{fl} + \mathbf{n}_\text{fl},
\label{FL_matrix}
\end{equation}
and the following SE analysis of FL is also based on (\ref{FL_matrix}).

\section{SE Analysis Framework}
\begin{figure*}[t]
\small
\begin{equation}
\text{SINR}_{\text{ml},n} = \frac{\frac{\rho_\text{ml}}{N_\text{t}}\left| E_{\mathbf{h}}\left\{ \mathbf{g}_{\text{ml},n}^H \mathbf{h}_{\text{ml},n} \right\} \right|^2} {\sum\limits_{n'=1}^{N_\text{t}} \frac{\rho_\text{ml}}{N_\text{t}} E_{\mathbf{h}}\left\{ \left| \mathbf{g}_{\text{ml},n}^H \mathbf{h}_{\text{ml},n'}\right|^2 \right\} - \frac{\rho_\text{ml}}{N_\text{t}} \left| E_{\mathbf{h}}\left\{ \mathbf{g}_{\text{ml},n}^H \mathbf{h}_{\text{ml},n} \right\} \right|^2 + \sum\limits_{m=1}^{N_\text{t}} \frac{\rho_\text{fl}}{N_\text{t}} E_{\mathbf{h}}\left\{ \left| \mathbf{g}_{\text{ml},n}^H \mathbf{h}_{\text{ml},m}\right|^2 \right\} + \sigma_\text{ml}^{2}E_{\mathbf{h}}\left\{ \left\| \mathbf{g}_{\text{ml},n} \right\|^2 \right\}},
\label{SINR_ML}
\end{equation}
\begin{equation}
I\left(\hat{\mathbf{y}}_{\text{ml}};\mathbf{x}_{\text{ml}}\right) = \log_2(N_\text{t}) - N_\text{t} + \frac{1}{N_\text{t}} \left\{ \sum_{n=1}^{N_\text{t}} \log_2\left(1+N_\text{t}\text{SINR}_{\text{ml},n}\right) - \sum_{n = 1}^{N_\text{t}} \log_2\left[\sum_{n'=1}^{N_\text{t}} \frac{\det\left(\mathbf{\Sigma}_{\text{ml},n}\right)} {\det\left(\mathbf{\Sigma}_{\text{ml},n} + \mathbf{\Sigma}_{\text{ml},n'}\right)}\right] \right\},
\label{SE_ML}
\end{equation}
\hrulefill
\end{figure*}

In this section, the SE analysis framework for SM-LDM systems with linear combining is separately proposed for ML and FL. Besides, the SE analysis frameworks for single-TA LDM systems and SMX aided LDM (SMX-LDM) systems are also proposed. Moreover, our proposed SE analysis framework can be easily extended to the multi-layer SM-LDM systems.

\subsection{Analysis for ML}

The received symbol of ML in (\ref{ML_matrix}) can be transformed as a vector form, which is denoted as follows:
\begin{equation}
\begin{split}
\mathbf{y}_\text{ml} &= \sum_{n=1}^{N_\text{t}} \sqrt{\rho_\text{ml}} s_{\text{ml},n} \gamma_{\text{ml},n} \mathbf{h}_{\text{ml},n} \\
&+ \sum_{m=1}^{N_\text{t}} \sqrt{\rho_\text{fl}} s_{\text{fl},m} \gamma_{\text{fl},m} \mathbf{h}_{\text{ml},m} + \mathbf{n}_\text{ml},
\label{ML_vector}
\end{split}
\end{equation}
where $s_{\text{ml},n} \sim \mathcal{CN}(0,1)$ and $s_{\text{fl},m} \sim \mathcal{CN}(0,1)$ denote the Gaussian inputs of ML and FL, respectively. $\gamma_{\text{ml},n}$ and $\gamma_{\text{fl},m}$ represent the activity of the $n$-th TA for ML and $m$-th TA for FL, respectively. Aided by SM property, $\sum_{n=1}^{N_\text{t}} \gamma_{\text{ml},n} = \sum_{m=1}^{N_\text{t}} \gamma_{\text{fl},m} = 1$, $\mathcal{P}(\gamma_{\text{ml},n} = 1) = \mathcal{P}(\gamma_{\text{fl},m} = 1) = \frac{1}{N_\text{t}}$ and $\mathcal{P}(\gamma_{\text{ml},n} = 0) = \mathcal{P}(\gamma_{\text{fl},m} = 0) = \frac{N_\text{t}-1}{N_\text{t}}$. In addition, $\mathbf{h}_{\text{ml},n} \in \mathbb{C}^{N_\text{rm}\times 1}$ denotes the $n$-th column of $\mathbf{H}_\text{ml}$.

With linear combining, $\mathbf{g}_{\text{ml},n}$ is denoted as the combining vector for the $n$-th TA in ML. Therefore, the SINR corresponding to the $n$-th TA of ML, i.e., $\text{SINR}_{\text{ml},n}$ can be lower bounded as (\ref{SINR_ML}), which can be proved from a direct application of \cite[Lemma~1]{Massive_SINR}. In (\ref{SINR_ML}), the numerator denotes the received power of the needed $n$-th transmit symbol in ML. The first two terms of the denominator in (\ref{SINR_ML}) represent the received power of other transmit symbols in ML, i.e., the inter-symbol-interference (ISI) introduced by ML. The third term of the denominator in (\ref{SINR_ML}) denotes the received power of transmit symbols in FL, which can be regarded as the ISI introduced by FL. The forth term of the denominator in (\ref{SINR_ML}) represents the influence of AWGN. Besides, the abbreviation $E_{\mathbf{h}}\{\cdot\}$  represents taking expectations over random realizations of channel vector $\mathbf{h}$.

Aided by the SINR expression in (\ref{SINR_ML}), an additive noise approximation can be introduced to (\ref{ML_vector}), and (\ref{ML_vector}) can be transformed as follows:
\begin{equation}
\hat{\mathbf{y}}_{\text{ml}} = \mathbf{x}_{\text{ml}} + \mathbf{w}_{\text{ml}},
\label{ML_equal}
\end{equation}
where $\hat{\mathbf{y}}_{\text{ml}} \in \mathbb{C}^{N_\text{t}\times 1}$ represents the equivalent received symbol in ML, and $\mathbf{w}_\text{ml}\in \mathbb{C}^{N_\text{t}\times 1}$ is a circularly symmetric complex-valued Gaussian noise, whose mean is $\mathbf{0}$ and the covariance matrix is denoted as follows:
\begin{equation}
E\left\{\mathbf{w}_\text{ml}\mathbf{w}_\text{ml}^H\right\} = \text{diag}\left\{\frac{1}{\text{SINR}_{\text{ml},1}},\ldots,\frac{1}{\text{SINR}_{\text{ml},N_\text{t}}}\right\}.
\end{equation}

Thus the MI can be divided into the spatial-domain MI and constellation-domain MI, which can be denoted as follows:
\begin{equation}
I\left(\hat{\mathbf{y}}_{\text{ml}};\mathbf{x}_{\text{ml}}\right)= I\left(\hat{\mathbf{y}}_{\text{ml}};\mathbf{a}_\text{ml}\right) + I\left(\hat{\mathbf{y}}_{\text{ml}};\mathbf{x}_{\text{ml}}|\mathbf{a}_\text{ml}\right).
\label{ML_split}
\end{equation}

Then aided by SM principle, in SM-LDM systems, the SE of ML with linear combining can be derived, and Theorem \ref{theorem_1} is introduced.

\begin{theorem}
The downlink SE of ML in SM-LDM systems with linear combining can be lower bounded as (\ref{SE_ML}), where $\mathbf{\Sigma}_{\text{ml},n}$ can be denoted as follows:
\begin{equation}
\mathbf{\Sigma}_{\text{ml},n} = \text{diag}\left\{\frac{1}{\text{SINR}_{\text{ml},1}}, ..., \frac{1}{\text{SINR}_{\text{ml},N_\text{t}}}\right\} + N_\text{t} \text{diag}\{\hat{\mathbf{a}}_{\text{ml},n}\},
\label{Sigma}
\end{equation}
and $\hat{\mathbf{a}}_{\text{ml},n}$ represents the $n$-th column of an $N_\text{t}$-by-$N_\text{t}$ identity matrix $\mathbf{I}_{N_\text{t}}$.
\label{theorem_1}
\end{theorem}

\begin{IEEEproof}
When the active antenna of ML is determined, the constellation-domain MI in (\ref{ML_split}) can be quantified by Shannon's continuous-input continuous-output channel (CMCC) capacity \cite{yang2008information}, and thus we have:
\begin{equation}
I(\hat{\mathbf{y}}_{\text{ml}};\mathbf{x}_{\text{ml}}|\mathbf{a}_\text{ml}) = \frac{1}{N_\text{t}} \sum_{n=1}^{N_\text{t}} \log_2(1+N_\text{t}\text{SINR}_{\text{ml},n}).
\label{CMCC}
\end{equation}

According to the definition of MI, the spatial-domain MI term in (\ref{ML_split}) can be denoted as follows:
\begin{equation}
\begin{split}
&I(\hat{\mathbf{y}}_{\text{ml}};\mathbf{a}_\text{ml}) = \text{T}_1 - \text{T}_2  \\
&=\frac{1}{N_\text{t}} \int \sum_{n=1}^{N_\text{t}} \mathcal{P} (\hat{\mathbf{y}}_{\text{ml}}|\hat{\mathbf{a}}_{\text{ml},n}) \log_2 \mathcal{P} (\hat{\mathbf{y}}_{\text{ml}}|\hat{\mathbf{a}}_{\text{ml},n}) d\hat{\mathbf{y}}_{\text{ml}} - \\
&\frac{1}{N_\text{t}} \int \sum_{n=1}^{N_\text{t}} \mathcal{P} (\hat{\mathbf{y}}_{\text{ml}}|\hat{\mathbf{a}}_{\text{ml},n}) \log_2 \left[\frac{1}{N_\text{t}} \sum_{n'=1}^{N_\text{t}} \mathcal{P} (\hat{\mathbf{y}}_{\text{ml}}|\hat{\mathbf{a}}_{\text{ml},n'})\right] d\hat{\mathbf{y}}_{\text{ml}},
\label{ML_AD_split}
\end{split}
\end{equation}
where $\mathcal{P} (\hat{\mathbf{y}}_{\text{ml}}|\hat{\mathbf{a}}_{\text{ml},n}) = \mathcal{CN}(\hat{\mathbf{y}}_{\text{ml}} ; \mathbf{0}, \mathbf{\Sigma}_{\text{ml},n})$ is a likelihood function.

In (\ref{ML_AD_split}), the term $\text{T}_1$ can be directly calculated out as follows:
\begin{equation}
\begin{array}{ll}
\text{T}_1 = -N_\text{t}\log_2(\pi e) -  \frac{1}{N_\text{t}}\sum_{n=1}^{N_\text{t}} \log_2(\det(\mathbf{\Sigma}_{\text{ml},n})).
\label{T1}
\end{array}
\end{equation}

However, the term $\text{T}_2$ lacks a closed-form solution, so the Jensen's inequality is introduced for approximation as follows:
\begin{equation}
\begin{split}
&\text{T}_2 \leq \\
&\frac{1}{N_\text{t}} \sum_{n=1}^{N_\text{t}} \log_2 \left[\frac{1}{N_\text{t}} \sum_{n'=1}^{N_\text{t}} \int \mathcal{P} (\hat{\mathbf{y}}_{\text{ml}}|\hat{\mathbf{a}}_{\text{ml},n}) \mathcal{P} (\hat{\mathbf{y}}_{\text{ml}}|\hat{\mathbf{a}}_{\text{ml},n'}) d\hat{\mathbf{y}}_{\text{ml}} \right] \\
& = \frac{1}{N_\text{t}} \sum_{n=1}^{N_\text{t}} \log_2\left[  \sum_{n'=1}^{N_\text{t}} \frac{\frac{1}{N_\text{t}}} {\det(\mathbf{\Sigma}_{\text{ml},n} + \mathbf{\Sigma}_{\text{ml},n'})} \right] - N_\text{t}\log_2\pi.
\end{split}
\label{T2}
\end{equation}

\begin{figure*}[t]
\small
\begin{equation}
\text{SINR}_{\text{fl},m} = \frac{\frac{\rho_\text{fl}}{N_\text{t}} \left| E_{\mathbf{h}}\left\{ \mathbf{g}_{\text{fl},m}^H \mathbf{h}_{\text{fl},m} \right\} \right|^2} {\sum_{m'=1}^{N_\text{t}} \frac{\rho_\text{fl}}{N_\text{t}} E_{\mathbf{h}}\left\{ \left| \mathbf{g}_{\text{fl},m}^H \mathbf{h}_{\text{fl},m'}\right|^2 \right\} - \frac{\rho_\text{fl}}{N_\text{t}} \left| E_{\mathbf{h}}\left\{ \mathbf{g}_{\text{fl},m}^H \mathbf{h}_{\text{fl},m} \right\} \right|^2 + \sigma_\text{fl}^{2}E_{\mathbf{h}}\left\{ \left\| \mathbf{g}_{\text{fl},m} \right\|^2 \right\}},
\label{SINR_FL}
\end{equation}
\begin{equation}
I\left(\hat{\mathbf{y}}_{\text{fl}};\mathbf{x}_{\text{fl}}\right) = \log_2\left(N_\text{t}\right) - N_\text{t} + \frac{1}{N_\text{t}} \left\{ \sum_{m=1}^{N_\text{t}} \log_2\left(1+N_\text{t}\text{SINR}_{\text{fl},m}\right) - \sum_{m = 1}^{N_\text{t}} \log_2\left[\sum_{m'=1}^{N_\text{t}} \frac{\det\left(\mathbf{\Sigma}_{\text{fl},m}\right)} {\det\left(\mathbf{\Sigma}_{\text{fl},m} + \mathbf{\Sigma}_{\text{fl},m'}\right)}\right] \right\},
\label{SE_FL}
\end{equation}
\hrulefill
\end{figure*}

By substituting (\ref{T1}) and (\ref{T2}) into (\ref{ML_AD_split}), the spatial-domain MI term can be lower bounded as follows:
\begin{equation}
\begin{split}
&I(\hat{\mathbf{y}}_{\text{ml}};\mathbf{a}_\text{ml}) \geq \log_2 N_\text{t} \\
&- \frac{1}{N_\text{t}} \sum_{n = 1}^{N_\text{t}} \log_2\left[\sum_{n'=1}^{N_\text{t}} \frac{\det(\mathbf{\Sigma}_{\text{ml},n})} {\det(\mathbf{\Sigma}_{\text{ml},n} + \mathbf{\Sigma}_{\text{ml},n'})}\right] - N_\text{t}\log_2 e.
\end{split}
\label{SD_AP}
\end{equation}

Moreover, aided by SM principle, when all SINRs of ML approximate to infinity, the spatial-domain MI of ML should approximate to $\log_2 N_\text{t}$. Besides, when all SINRs of ML approximate to $0$, the spatial-domain MI of ML should approximate to $0$. However, the limitations of derived lower bound in (\ref{SD_AP}) are different, and each limitation lacks a constant biase. To achieve an unbiased SE lower bound, a constant shift is applied in (\ref{SD_AP}), and the asymptotically unbiased spatial-domain MI lower bound can be derived as follows:
\begin{equation}
\begin{split}
&I(\hat{\mathbf{y}}_{\text{ml}};\mathbf{a}_\text{ml}) \succeq \log_2(N_\text{t}) - N_\text{t} \\
&- \frac{1}{N_\text{t}} \sum_{n = 1}^{N_\text{t}} \log_2 \left[\sum_{n'=1}^{N_\text{t}} \frac{\det(\mathbf{\Sigma}_{\text{ml},n})} {\det(\mathbf{\Sigma}_{\text{ml},n}+\mathbf{\Sigma}_{\text{ml},n'})} \right].
\end{split}
\label{SD}
\end{equation}

Therefore, by substituting (\ref{CMCC}) and (\ref{SD}) into (\ref{ML_split}), the SE lower bound of ML can be formulated as (\ref{SE_ML}), which completes this proof.
\end{IEEEproof}

From (\ref{SE_ML}), it can be illustrated that the SE of ML is composed of the constellation-domain part and the spatial-domain part, and both parts are increased with the increasing of SINR. Therefore, a higher SINR leads to a larger SE of ML. With a specific linear combining algorithm, the closed-form SINR can be derived, and then aided by Theorem \ref{theorem_1}, the theoretical value of SE in ML can be formulated.

\subsection{Analysis for FL}

Aided by (\ref{FL_matrix}), the received symbol of FL can also be transformed as a vector form as follows:
\begin{equation}
\mathbf{y}_\text{fl} = \sum_{m=1}^{N_\text{t}} \sqrt{\rho_\text{fl}} s_{\text{fl},m} \gamma_{\text{fl},m} \mathbf{h}_{\text{fl},m} + \mathbf{n}_\text{fl}.
\label{FL_vector}
\end{equation}

For FL, $\mathbf{g}_{\text{fl},m}$ represents the linear combining vector for the $m$-th TA, and the SINR of the $m$-th TA can be lowered bounded as (\ref{SINR_FL}). From (\ref{SINR_FL}), it can be seen that the numerator represents the received power of the transmit symbol of the $m$-th TA in FL, the first two terms of the denominator denote the ISI introduced by other transmit symbols in FL, and the last term of the denominator denotes the influence of AWGN. Different from the SINR of ML in (\ref{SINR_ML}), for SINR of FL, only the transmit symbols of FL introduce the ISI, and the transmit symbols of ML have no influence on the SINR of FL assuming perfect cancellation.

With respect to the SE lower bound of FL, from a direct application of the SE analysis for ML, Theorem \ref{theorem_2} can be introduced based on (\ref{FL_matrix}).
\begin{theorem}
The downlink SE of FL in SM-LDM systems with linear combining can be lower bounded as (\ref{SE_FL}), where $\mathbf{\Sigma}_{\text{fl},m}$ can be denoted as follows:
\begin{equation}
\mathbf{\Sigma}_{\text{fl},m} = \text{diag}\left\{\frac{1}{\text{SINR}_{\text{fl},1}}, ..., \frac{1}{\text{SINR}_{\text{fl},N_\text{t}}}\right\} + N_\text{t} \text{diag}\{\hat{\mathbf{a}}_{\text{fl},m}\},
\label{Sigma_FL}
\end{equation}
and $\hat{\mathbf{a}}_{\text{fl},m}$ denotes the $m$-th column of an $N_\text{t}$-by-$N_\text{t}$ identity matrix $\mathbf{I}_{N_\text{t}}$.
\label{theorem_2}
\end{theorem}

\begin{IEEEproof}
The proof of Theorem \ref{theorem_2} follows from a direct application of the proof of Theorem \ref{theorem_1}.
\end{IEEEproof}

From Theorem \ref{theorem_1} and Theorem \ref{theorem_2}, it can be shown that SE lower bound expressions for both ML and FL are almost the same, but the SINR of ML differs from the SINR of FL. More specifically, the SE of ML is influenced from both transmit symbols in FL and transmit symbols in ML, but the SE of FL is only influenced by transmit symbols in FL. This is because the perfect ML signal cancellation and perfect CE are assumed in this paper, and the CLI is not explicitly considered.

\subsection{Analysis for Single-TA LDM and SMX-LDM}
For conventional single-TA LDM systems, the SE of ML can be obtained by substituting $N_\text{t} = 1$ into (\ref{SE_ML}), and the SE of FL can be obtained by substituting $N_\text{t} = 1$ into (\ref{SE_FL}). Thus the SE of both ML and FL for single-TA LDM systems can be derived as follows:
\begin{equation}
R_\text{ml}^\text{ST} = \log_2\left(1+\text{SINR}_\text{ml}^\text{ST}\right), \ \
R_\text{fl}^\text{ST} = \log_2\left(1+\text{SINR}_\text{fl}^\text{ST}\right),
\label{SE_ST}
\end{equation}
where $R_\text{ml}^\text{ST}$ and $R_\text{fl}^\text{ST}$ represent the SE of ML and FL in single-TA LDM systems, respectively. In addition, $\text{SINR}_\text{ml}^\text{ST}$ and $\text{SINR}_\text{fl}^\text{ST}$ denote the SINR of ML and FL in single-TA LDM systems, respectively. The $\text{SINR}_\text{ml}^\text{ST}$ and $\text{SINR}_\text{fl}^\text{ST}$ can be obtained by substituting $N_\text{t} = 1$ into (\ref{SINR_ML}) and (\ref{SINR_FL}), respectively. In single-TA LDM systems, only the constellation symbols transmit information, so (\ref{SE_ST}) represents the exact value of SE. The approximation is only conducted when deriving the spatial-domain MI.

For SMX-LDM systems, since all transmit antennas are active to transmit constellation symbols, the SE of ML and FL can be quantified by CMCC capacity as follows:
\begin{equation}
\begin{split}
&R_\text{ml}^\text{SMX} = \sum_{n=1}^{N_\text{t}} \log_2\left(1+\text{SINR}_{\text{ml},n}^\text{SMX}\right), \\
&R_\text{fl}^\text{SMX} = \sum_{m=1}^{N_\text{t}} \log_2\left(1+\text{SINR}_{\text{fl},m}^\text{SMX}\right),
\end{split}
\label{SE_SMX}
\end{equation}
where $R_\text{ml}^\text{SMX}$ and $R_\text{fl}^\text{SMX}$ denote the SE of ML and FL in SMX-LDM systems, respectively. Besides, $\text{SINR}_{\text{ml},n}^\text{SMX}$ represents the SINR of the $n$-th TA in ML of SMX-LDM systems, and $\text{SINR}_{\text{fl},m}^\text{SMX}$ represents the SINR of the $m$-th TA in FL of SMX-LDM systems. Similarly, for SMX-LDM systems, (\ref{SE_SMX}) is the exact value rather than the lower bound, since only the constellation domain transmits information.

\section{Closed-Form SE Lower Bound with MRC}

In our proposed SE analysis framework, the SINR values are related to specific combining algorithms. In this section, MRC is considered for SM-LDM systems, single-TA LDM systems and SMX-LDM systems. In addition, the closed-form SE lower bound for SM-TDM/FDM systems with MRC is also formulated.

\subsection{SM-LDM}
In this subsection, MRC is considered for both ML and FL, and the SINR values of these two layers are derived as closed forms. Then substituting the closed-form SINR values into Theorem \ref{theorem_1} and Theorem \ref{theorem_2}, the closed-form SE lower bound of SM-LDM systems with MRC can be formulated.

For MRC, the combining vector of the $n$-th TA for ML is the estimated $n$-th column of $\mathbf{H}_\text{ml}$, and the combining vector of the $m$-th TA for FL is the estimated $m$-th column of $\mathbf{H}_\text{fl}$. Since the perfect CE is assumed, we have:
\begin{equation}
\mathbf{g}_{\text{ml},n} = \mathbf{h}_{\text{ml},n}, \ \ \mathbf{g}_{\text{fl},m} = \mathbf{h}_{\text{fl},m}.
\label{MRC_g}
\end{equation}

Aided by (\ref{MRC_g}), it can be immediately formulated as follows:
\begin{equation}
\begin{split}
&E_{\mathbf{h}}\left\{ \mathbf{g}_{\text{ml},n}^H\mathbf{h}_{\text{ml},n} \right\} = E_{\mathbf{h}}\left\{ \| \mathbf{g}_{\text{ml},n} \|^2 \right\} = N_\text{rm}, \\
&E_{\mathbf{h}}\left\{ \mathbf{g}_{\text{fl},m}^H\mathbf{h}_{\text{fl},m} \right\} = E_{\mathbf{h}}\left\{ \| \mathbf{g}_{\text{fl},m} \|^2 \right\} = N_\text{rf}.
\end{split}
\label{numerator}
\end{equation}

For the ISI terms in ML, if $n'\neq n$, $\mathbf{g}_{\text{ml},n}$ and $\mathbf{h}_{\text{ml},n'}$ are independent. When $m \neq n$, $\mathbf{g}_{\text{ml},n}$ and $\mathbf{h}_{\text{ml},m}$ are also independent. Therefore, in these cases we have:
\begin{equation}
\begin{split}
&E_{\mathbf{h}}\left\{ \left| \mathbf{g}_{\text{ml},n}^H \mathbf{h}_{\text{ml},n'}\right|^2 \right\} = E_{\mathbf{h}}\left\{ \left\| \mathbf{g}_{\text{ml},n} \right\|^2 \right\} = N_\text{rm}, \\
&E_{\mathbf{h}}\left\{ \left| \mathbf{g}_{\text{ml},n}^H \mathbf{h}_{\text{ml},m}\right|^2 \right\} = E_{\mathbf{h}}\left\{ \left\| \mathbf{g}_{\text{ml},n} \right\|^2 \right\} = N_\text{rm}.
\end{split}
\label{idpdt}
\end{equation}

In addition, if $n' = n$, $\mathbf{g}_{\text{ml},n}$ and $\mathbf{h}_{\text{ml},n'}$ are correlated. When $m = n$, $\mathbf{g}_{\text{ml},n}$ and $\mathbf{h}_{\text{ml},m}$ are also correlated. In these cases, aided by the property of the central complex-valued Wishart distribution \cite{Wishart}, we have:
\begin{equation}
\begin{split}
&E_{\mathbf{h}}\left\{ \left| \mathbf{g}_{\text{ml},n}^H \mathbf{h}_{\text{ml},n'}\right|^2 \right\} = E_{\mathbf{h}}\left\{ \left| \mathbf{g}_{\text{ml},n}^H \mathbf{h}_{\text{ml},m}\right|^2 \right\} \\
&= E_{\mathbf{h}}\left\{ \left\| \mathbf{h}_{\text{ml},n} \right\|^4 \right\} = N_\text{rm}(N_\text{rm} + 1).
\end{split}
\label{ucrltd}
\end{equation}

By substituting (\ref{numerator}), (\ref{idpdt}) and (\ref{ucrltd}) into (\ref{SINR_ML}), the SINR of the $n$-th TA in ML with MRC can be formulated as follows:
\begin{equation}
\text{SINR}_{\text{ml},n} = \frac{\rho_\text{ml}N_\text{rm}} {\rho_\text{ml}N_\text{t} + \rho_\text{fl}(N_\text{t}+N_\text{rm}) + N_\text{t}\sigma_\text{ml}^2}.
\label{SINR_ML_T}
\end{equation}

In (\ref{SINR_ML_T}), the numerator denotes the power of the targeted received symbol of ML, the first term of the denominator denotes the ISI caused by other symbols of ML, the second term of the denominator denotes the ISI cause by symbols of FL, and the last term of the denominator represents the AWGN. In addition, from (\ref{SINR_ML_T}), it can be seen that increasing the number of RAs in ML or decreasing the number of TAs can bring a larger SINR for ML. Besides, although enlarging the transmit power of ML can also increase the SINR of ML, the SINR cannot increase indefinitely because of the ISI caused by symbols from both ML and FL.

Following from a similar application of SINR derivation in ML, the SINR corresponding to the $m$-th TA of FL with MRC can be derived too, which can be denoted as follows:
\begin{equation}
\text{SINR}_{\text{fl},m} = \frac{\rho_\text{fl}N_\text{rf}} {\rho_\text{fl}N_\text{t} + N_\text{t}\sigma_\text{fl}^2}.
\label{SINR_FL_T}
\end{equation}

In (\ref{SINR_FL_T}), the numerator represents the power of the targeted received symbol in FL, the first term and the second term of the denominator represent the ISI caused by other symbols in FL and AWGN, respectively. Similarly, increasing the transmit power of FL can enlarge the SINR of FL. More RAs in FL or less TAs can also increase the SINR of FL.

Aided by the SINR of ML in (\ref{SINR_ML_T}), the closed-form SE lower bound for ML in SM-LDM systems with MRC can be derived by substituting (\ref{SINR_ML_T}) into (\ref{SINR_ML}) and (\ref{SE_ML}). The closed-form SE lower bound for FL in SM-LDM systems with MRC can also be derived by substituting (\ref{SINR_FL_T}) into (\ref{SINR_FL}) and (\ref{SE_FL}).

\subsection{Single-TA LDM and SMX-LDM}

For single-TA LDM systems, the SINR of ML and FL can be derived by applying $N_\text{t} = 1$ into (\ref{SINR_ML_T}) and (\ref{SINR_FL_T}), respectively. Thus we have:
\begin{equation}
\begin{split}
&\text{SINR}_\text{ml}^\text{ST} = \frac{\rho_\text{ml}N_\text{rm}} {\rho_\text{ml}+ \rho_\text{fl}(1 + N_\text{rm}) + \sigma_\text{ml}^2},\\
&\text{SINR}_\text{fl}^\text{ST} = \frac{\rho_\text{fl}N_\text{rf}} {\rho_\text{fl} + \sigma_\text{fl}^2}.
\end{split}
\label{SINR_ST}
\end{equation}

By substituting (\ref{SINR_ST}) into (\ref{SE_ST}), the SE exact value of single-TA LDM systems with MRC is derived. Comparing (\ref{SINR_ST}) with (\ref{SINR_ML_T}) and (\ref{SINR_FL_T}), the SINR of ML and FL in single-TA LDM systems is larger than the SINR of ML and FL for SM-LDM systems, respectively. Therefore, the constellation-domain MI of single-TA LDM systems is larger than that of SM-LDM systems. However, the spatial domain can also be utilized for information transmission in SM-LDM systems, so the SE comparison between SM-LDM systems and single-TA LDM systems is conducted in the section of simulation results.

For SMX-LDM systems, to ensure the fairness of the same transmit power, comparing with SM-LDM systems, the transmit power of each TA should divide $N_\text{t}$. Thus we have:
\begin{equation}
\begin{split}
&\text{SINR}_{\text{ml},n}^\text{SMX} = \frac{\rho_\text{ml}N_\text{rm}} {\rho_\text{ml}N_\text{t} + \rho_\text{fl}(N_\text{t}+N_\text{rm}) + N_\text{t}^2\sigma_\text{ml}^2}, \\
&\text{SINR}_{\text{fl},m}^\text{SMX} = \frac{\rho_\text{fl}N_\text{rf}} {\rho_\text{fl}N_\text{t} + N_\text{t}^2\sigma_\text{fl}^2}.
\end{split}
\label{SINR_SMX}
\end{equation}

\begin{figure*}[t]
\small
\begin{equation}
S^\text{TF}_\text{ml} = \frac{L_\text{ml}}{L_\text{ml} + L_\text{fl}}\left\{ \log_2(N_\text{t}) - N_\text{t} + \frac{1}{N_\text{t}} \left[ \sum_{n=1}^{N_\text{t}} \log_2\left(1+N_\text{t}\text{SINR}^\text{TF}_{\text{ml},n}\right) - \sum_{n = 1}^{N_\text{t}} \log_2\left(\sum_{n'=1}^{N_\text{t}} \frac{\det\left(\mathbf{\Sigma}^\text{TF}_{\text{ml},n}\right)} {\det\left(\mathbf{\Sigma}^\text{TF}_{\text{ml},n} + \mathbf{\Sigma}^\text{TF}_{\text{ml},n'}\right)}\right) \right] \right\},
\label{SE_ML_TFDM}
\end{equation}
\begin{equation}
S^\text{TF}_\text{fl} = \frac{L_\text{fl}}{L_\text{ml} + L_\text{fl}}\left\{ \log_2\left(N_\text{t}\right) - N_\text{t} + \frac{1}{N_\text{t}} \left[ \sum_{m=1}^{N_\text{t}} \log_2\left(1+N_\text{t}\text{SINR}^\text{TF}_{\text{fl},m}\right) - \sum_{m = 1}^{N_\text{t}} \log_2\left(\sum_{m'=1}^{N_\text{t}} \frac{\det\left(\mathbf{\Sigma}^\text{TF}_{\text{fl},m}\right)} {\det\left(\mathbf{\Sigma}^\text{TF}_{\text{fl},m} + \mathbf{\Sigma}^\text{TF}_{\text{fl},m'}\right)}\right) \right] \right\},
\label{SE_FL_TFDM}
\end{equation}
\hrulefill
\end{figure*}

By substituting (\ref{SINR_SMX}) into (\ref{SE_SMX}), the SE exact value of SMX-LDM systems with MRC can also be formulated. From (\ref{SINR_SMX}), (\ref{SINR_ML_T}) and (\ref{SINR_FL_T}), it can be seen that for SM-LDM systems and SMX-LDM systems, the ISI terms of both ML and FL have a same influence on the SINR of ML and FL, respectively. However, the AWGN terms of both ML and FL in SMX-LDM systems are larger than those in SM-LDM systems, which is because the transmit power of each TA in SMX-LDM systems is smaller than that in SM-LDM systems.

\subsection{SM-TDM/FDM}
For SM-TDM/FDM systems, the ML services and FL services are transmitted separately in time domain or frequency domain. Therefore, in SM-TDM/FDM systems we have $\rho_\text{ml}^\text{TF} = \rho_\text{fl}^\text{TF} = P_\text{u}$, where $\rho_\text{ml}^\text{TF}$ and $\rho_\text{fl}^\text{TF}$ represent the transmit power of ML and FL in SM-TDM/FDM systems, respectively. Then following from a same analysis of Section \uppercase\expandafter{\romannumeral3} A, the SE of ML and FL in SM-TDM/FDM systems can be lower bounded as (\ref{SE_ML_TFDM}) and (\ref{SE_FL_TFDM}), where $S^\text{TF}_\text{ml}$ and $S^\text{TF}_\text{fl}$ denote the SE lower bound of ML and FL in SM-TDM/FDM systems, respectively. The SINR of the $n$-th TA for ML and the SINR of the $m$-th TA for FL in SM-TDM/FDM systems are denoted as $\text{SINR}^\text{TF}_{\text{ml},n}$ and $\text{SINR}^\text{TF}_{\text{fl},m}$, respectively. With respect to $L_\text{ml}$ and $L_\text{fl}$, for SM-TDM systems, $L_\text{ml} + L_\text{fl}$ denotes the total time duration, and $L_\text{ml}$ and $L_\text{fl}$ are transmission time for ML and FL, respectively. For SM-FDM systems, $L_\text{ml} + L_\text{fl}$ represent the total bandwidth, and $L_\text{ml}$ and $L_\text{fl}$ are bandwidth for ML and FL, respectively. In addition, $\mathbf{\Sigma}^\text{TF}_{\text{ml},n}$ can be denoted as follows:
\begin{equation}
\mathbf{\Sigma}^\text{TF}_{\text{ml},n} = \text{diag}\left\{\frac{1}{\text{SINR}^\text{TF}_{\text{ml},1}}, ..., \frac{1}{\text{SINR}^\text{TF}_{\text{ml},N_\text{t}}}\right\} + N_\text{t} \text{diag}\{\hat{\mathbf{a}}_{\text{ml},n}\},
\label{Sigma_TF_ML}
\end{equation}
and $\mathbf{\Sigma}^\text{TF}_{\text{fl},m}$ can be denoted as follows:
\begin{equation}
\mathbf{\Sigma}^\text{TF}_{\text{fl},m} = \text{diag}\left\{\frac{1}{\text{SINR}^\text{TF}_{\text{fl},1}}, ..., \frac{1}{\text{SINR}^\text{TF}_{\text{fl},N_\text{t}}}\right\} + N_\text{t} \text{diag}\{\hat{\mathbf{a}}_{\text{fl},m}\}.
\label{Sigma_TF_FL}
\end{equation}

Since ML services and FL services are transmitted separately in SM-TDM/FDM systems, only ML transmit symbols constitute the ISI of the $\text{SINR}^\text{TF}_{\text{ml},n}$, and only FL transmit symbols constitute the ISI of the $\text{SINR}^\text{TF}_{\text{fl},m}$. Besides, for both transmission of ML and FL, the transmit power need not to be split. Therefore, it can be easily derived as follows:
\begin{equation}
\begin{split}
&\text{SINR}^\text{TF}_{\text{ml},n} = \frac{P_\text{u}N_\text{rm}} { N_\text{t}(P_\text{u} + \sigma_\text{ml}^2)}, \\
&\text{SINR}^\text{TF}_{\text{fl},m} = \frac{P_\text{u}N_\text{rf}} { N_\text{t}(P_\text{u} + \sigma_\text{fl}^2)},
\label{SINR_TF}
\end{split}
\end{equation}
and thus the closed-form SE lower bound for both ML and FL in SM-TDM/FDM systems can also be formulated.

\section{Simulation Results}
In this section, Monte Carlo simulations are provided to verify the tightness of the SE lower bound for SM-LDM systems, and the SE comparison between SM-LDM systems, single-TA LDM systems and SMX-LDM systems is also illustrated via simulations. Besides, we set $\text{SNR}$ rather than $\text{SINR}$ as the x-coordinate in Figures \ref{Simulation_SNR}-\ref{VB_SNRf_Nt}, which is because from Section \uppercase\expandafter{\romannumeral4}, it can be demonstrated that the $\text{SINR}$ is an intermediate variable depending on the number of TAs, the number of RAs and $\text{SNR}$. Thus using the independent variable $\text{SNR}$ as that in \cite{Massive_SINR} is more reasonable.

\subsection{Bound Tightness}
\begin{figure}
  \centering
  \includegraphics[width=0.45\textwidth]{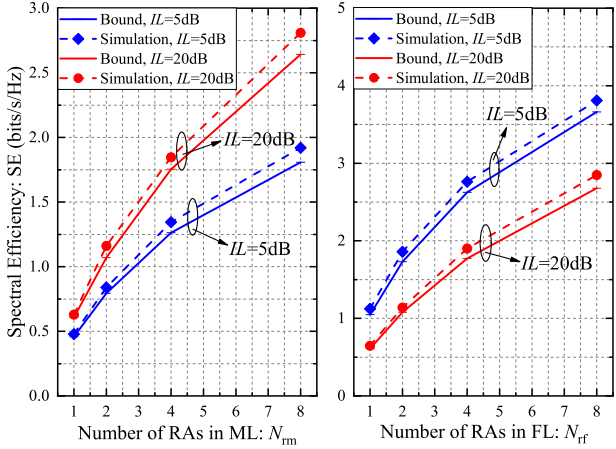}
  \caption{The SE performance of simulation results and our proposed SE lower bound versus $N_\text{rm}$ for ML in (a). The SE performance of simulation results and our proposed SE lower bound versus $N_\text{rf}$ for FL in (b).}\label{Simulation_Nr}
\end{figure}

In this subsection, the tightness of our proposed SE lower bound for SM-LDM systems is verified. In Fig. \ref{Simulation_Nr} (a), $N_\text{t} = 2$, $N_\text{rf} = 2$, $N_\text{rm} \in \{1,2,4,8\}$, $\textit{IL} \in \{5 \text{ dB}, 20 \text{ dB}\}$ and $\text{SNR}_\text{ml} = 0 \text{ dB}$ are assumed, where $\text{SNR}_\text{ml}$ denotes the SNR of ML. In Fig. \ref{Simulation_Nr} (b), we assume $N_\text{t} = 2$, $N_\text{rm} = 2$, $N_\text{rf} \in \{1,2,4,8\}$, $\textit{IL} \in \{5 \text{ dB}, 20 \text{ dB}\}$ and $\text{SNR}_\text{fl} = 0 \text{ dB}$, where $\text{SNR}_\text{fl}$ denotes the SNR of FL. As shown in Fig. \ref{Simulation_Nr}, our proposed SE lower bound is relatively tight. In addition, for both ML and FL, a larger number of RAs brings a larger gap between our proposed SE lower bound and simulated SE. This is because more RAs brings a larger SE, but the proportion of SE lower bound and simulated SE almost remains unchanged. As the growing of RAs, although the gap between SE lower bound and simulated SE becomes slightly bigger, the SE lower bound and SE simulation results also have the same trend.

\begin{figure}
  \centering
  \includegraphics[width=0.45\textwidth]{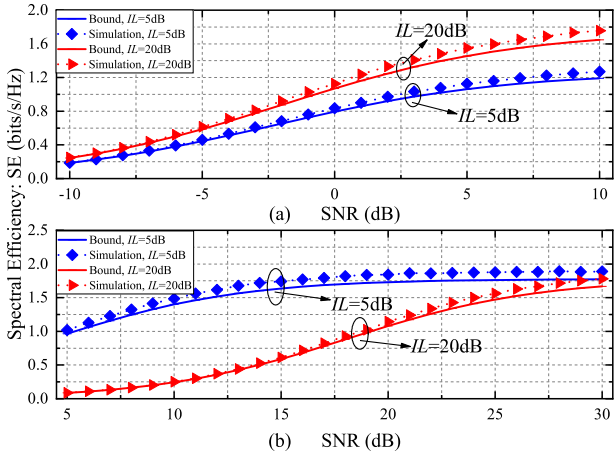}
  \caption{The SE performance versus SNR based on simulation results and our proposed SE lower bound for ML in (a) and for FL in (b).}\label{Simulation_SNR}
\end{figure}

In Fig. \ref{Simulation_SNR}, the system configurations include $N_\text{t} = 2$, $N_\text{rm} = 2$, $N_\text{rf} = 2$ and $\textit{IL} \in \{5 \text{ dB}, 20 \text{ dB}\}$. From Fig. \ref{Simulation_SNR} (a), it can be observed that a larger $\text{SNR}_\text{ml}$ leads to a higher SE in ML. From Fig. \ref{Simulation_SNR} (b), although a larger $\text{SNR}_\text{fl}$ can also bring a higher SE in FL, when $\text{SNR}_\text{fl}$ becomes relatively high, the SE of FL becomes almost saturated. This is because with quite high SNR, the ISI mainly brings influence on this interference-limited system. In addition, from both Fig. \ref{Simulation_Nr} and Fig. \ref{Simulation_SNR}, it can be illustrated that a larger \textit{IL} brings a higher SE of ML, and a lower SE of FL.

In a word, our proposed SE lower bound of SM-LDM systems are relatively tight, and the bound and simulation results is the same trend. Therefore, this SE lower bound will be utilized for the SE comparison in the next subsection.

\subsection{SE Comparison}

\begin{figure}
  \centering
  \includegraphics[width=0.45\textwidth]{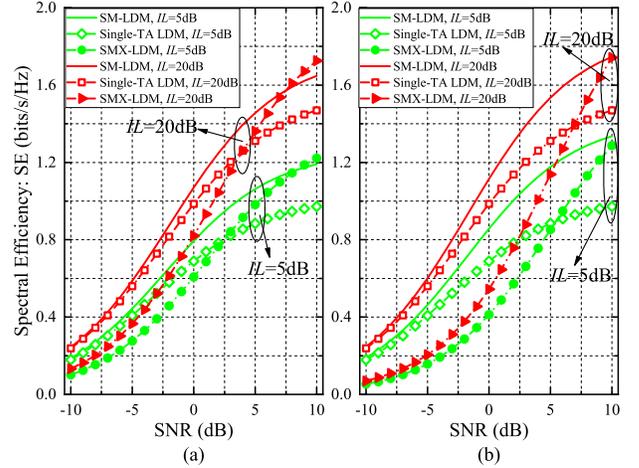}
  \caption{The SE performance of ML in SM-LDM, single-TA LDM and SMX-LDM systems versus SNR with $N_\text{rm} = 2$, $N_\text{rf} = 2$ and $\textit{IL} \in \{5 \text{ dB}, 20 \text{ dB}\}$. $N_\text{t} = 2$ in (a) and $N_\text{t} = 4$ in (b).}\label{VB_SNRm_Nt}
\end{figure}

In this subsection, the SE comparison between different schemes are proposed via simulations. It should be pointed out that, $N_\text{t}$ denotes the number of TAs in SM-LDM systems and SMX-LDM systems, but for single-TA LDM systems, we have $N_\text{t} = 1$. In addition, although in practice for Digital Terrestrial Television (DTT) the number of TAs in MIMO is $2$ \cite{ATSC_MIMO}, recently the MIMO systems with more than $2$ TAs, even massive MIMO systems have also been considered in broadcasting transmission scenarios \cite{SM_TBC} \cite{SM_Jiao}. Therefore, in this subsection, we set $N_\text{t} \in \{1,2,4\}$.

From Fig. \ref{VB_SNRm_Nt}, it can be observed that the SM-LDM system always has a higher ML SE than that of the single-TA LDM system, which is because the spatial domain transmits extra information. Additionally, our proposed SM-LDM system even has a better ML SE performance than that of the SMX-LDM system in low SNR region. For ML transmission, the SNR is relatively low, and in this case the AWGN mainly brings influence on this power-limited system. Comparing with (\ref{SINR_ML_T}), (\ref{SINR_FL_T}) and (\ref{SINR_SMX}), the SMX-LDM system has a much lower SINR than that of the SM-LDM system. In addition, from Fig. \ref{VB_SNRm_Nt} (a) and Fig. \ref{VB_SNRm_Nt} (b), it can be illustrated that increasing $N_\text{t}$ can also increase the ML SE in SM-LDM systems, which is because a larger $N_\text{t}$ brings a higher spatial-domain MI. However, for SMX-LDM systems, the SE with $N_\text{t} = 4$ is smaller than the SE with $N_\text{t} = 2$ when SNR is low. This is because for fairness, the transmit power of each TA in SMX-LDM systems divides $N_\text{t}$. From (\ref{SINR_SMX}) and (\ref{SINR_ML_T}), it can be seen that comparing with SM-LDM systems, the SE of SMX-LDM systems is much influenced by growing of $N_\text{t}$ in low SNR region. Since the SNR is always not high for ML, our proposed SM-LDM system is pretty suitable.

\begin{figure}
  \centering
  \includegraphics[width=0.45\textwidth]{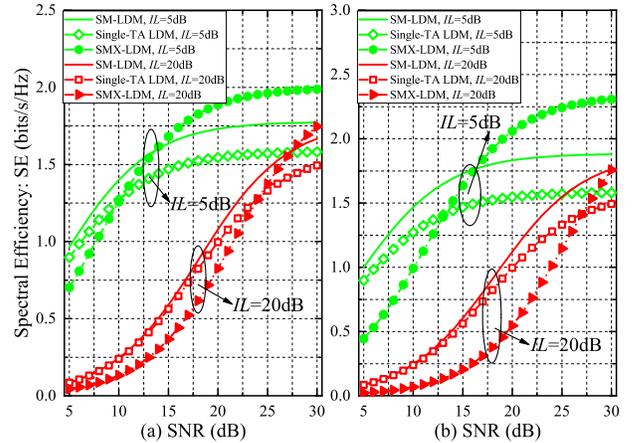}
  \caption{The SE performance of FL in SM-LDM, single-TA LDM and SMX-LDM systems versus SNR with $N_\text{rm} = 2$, $N_\text{rf} = 2$ and $\textit{IL} \in \{5 \text{ dB}, 20 \text{ dB}\}$. $N_\text{t} = 2$ in (a) and $N_\text{t} = 4$ in (b).}\label{VB_SNRf_Nt}
\end{figure}

As shown in Fig. \ref{VB_SNRf_Nt}, the SM-LDM system still has a higher FL SE than that of the single-TA LDM system because of the spatial-domain information. However, as the SNR becomes larger in FL, the FL SE of the SMX-LDM system exceeds the FL SE of the SM-LDM system. This is because with a relatively high SNR for FL transmission, the ISI rather than the AWGN mostly influences the SINR. From (\ref{SINR_SMX}) and (\ref{SINR_FL_T}), in FL, the ISI terms for both the SM-LDM systems and SMX-LDM systems are similar, and in high SNR region, the AWGN terms for SM-LDM systems and SMX-LDM systems almost have the same influence on SINR. Therefore, with a high SNR in FL, the SINR of SM-LDM systems are almost the same as the SINR of SMX-LDM systems. In this case the constellation-domain MI in SMX-LDM systems is higher than the spatial-domain MI in SM-LDM systems, so with a high SNR the FL SE of SMX-LDM systems is higher than the FL SE of SM-LDM systems. In addition, for SMX-LDM systems, in low SNR region, the FL SE with $N_\text{t} = 4$ is lower than that with $N_\text{t} = 2$, but in high SNR region, the FL SE with $N_\text{t} = 4$ is higher than that with $N_\text{t} = 2$. This is because in low SNR region, this system is a power-limited system, but in high SNR region, this system is an interference-limited system.

\begin{figure}
  \centering
  \includegraphics[width=0.45\textwidth]{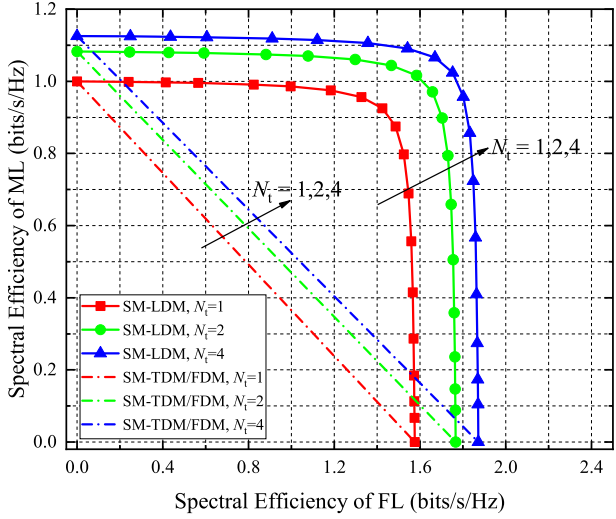}
  \caption{The SE performance of ML and FL in SM-LDM and SM-TDM/FDM systems with $N_\text{t} \in \{1,2,4\}$, $N_\text{rm} = 2$, $N_\text{rf} = 2$, $\text{SNR}_\text{ml} = 0 \text{ dB}$ and $\text{SNR}_\text{fl} = 20 \text{ dB}$.}\label{TFDM_SM_LDM}
\end{figure}

In Fig. \ref{TFDM_SM_LDM}, we compare the SE between SM-LDM systems and SM-TDM/FDM systems with different $N_\text{t}$. It can be observed that a larger $N_\text{t}$ leads to a higher SE, but the SE gain between $N_\text{t} = 4$ and $N_\text{t} = 2$ is lower than that between $N_\text{t} = 2$ and $N_\text{t} = 1$. This is because although increasing $N_\text{t}$ leads to a larger spatial-domain MI, a larger $N_\text{t}$ also brings a larger ISI in (\ref{SINR_ML_T}) and (\ref{SINR_FL_T}). It can also be illustrated that the SM-LDM systems can outperform the SE of SM-TDM/FDM systems. In addition, for FL in SM-LDM systems, when \textit{IL} is small enough, decreasing \textit{IL} can hardly increase the FL SE, which is because in a small \textit{IL}, it is a interference-limited system for FL. Similarly, when \textit{IL} is large enough, increasing \textit{IL} can also barely increase the SE of ML.

\section{Conclusion}
In this paper, a SM-LDM system is proposed to increase the SE for terrestrial broadcasting transmission. The SE analysis framework is proposed with linear combining algorithms, and the closed-form SE lower bound for SM-LDM systems with MRC are also derived. In addition, for comparison, the closed-form SE of traditional single-TA LDM systems and SMX-LDM systems is also formulated. Our proposed SE analysis scheme can also be easily extended to the multi-layer SM-LDM systems. Simulation results are provided to validate the tightness of our proposed SE lower bound for SM-LDM systems, and SM-LDM systems can outperform the SE of SM-TDM/FDM systems and single-TA LDM systems. The SM-LDM systems can even have a higher SE than SMX-LDM systems in low SNR region via simulations.

\ifCLASSOPTIONcaptionsoff
  \newpage
\fi

%





\begin{thebibliography}{1}

\bibitem{ATSC}
L. Fay, L. Michael, D. G\'{o}mez-Barquero, N. Ammar, and M. W. Caldwell, ``An overview of the ATSC 3.0 physical layer specification,'' \emph{IEEE Trans. Broadcast.}, vol. 62, no. 1, pp. 233-243, Mar. 2016.

\bibitem{ATSC_MIMO}
T. Shitomi, E. Garro, K. Murayama, and D. G\'{o}mez-Barquero, ``MIMO scattered pilot performance and optimization for ATSC 3.0,'' \emph{IEEE Trans. Broadcast.}, to be published.

\bibitem{LDM_ATSC}
S. I. Park, J. Y. Lee, S. Myoung, L. Zhang, Y. Wu, J. Montalb\'{a}n, S. Kwon, B. M. Lim, P. Angueira, H. M. Kim, N. Hur, and J. Kim, ``Low complexity layered division multiplexing system for ATSC 3.0,'' \emph{IEEE Trans. Broadcast.}, vol. 62, no. 1, pp. 233-243, Mar. 2016.

\bibitem{LDM_TAP}
L. Zhang, W. Li, Y. Wu, X. Wang, S. I. Park, H. M. Kim, J. Y. Lee, P. Angueira, and J. Montalban, ``Layered-division-multiplexing: Theory and practice,'' \emph{IEEE Trans. Broadcast.}, vol. 62, no. 1, pp. 216-232, Mar. 2016.

\bibitem{LDM_MIMO_C}
D. G\'{o}mez-Barquero and O. Simeone, ``LDM vs. FDM/TDM for unequal error protection in terrestrial broadcasting systems: An informationtheoretic view,'' \emph{IEEE Trans. Broadcast.}, vol. 61, no. 4, pp. 571-579, Dec. 2015.

\bibitem{LDM_BMSB}
L. Zhang, Y. Wu, W. Li, H. M. Kim, S. I. Park, P. Angueira, J. Montalban, and M. Velez, ``Channel capacity distribution of layer-division-multiplexing system for next generation digital broadcasting transmission,'' in \emph{Proc. IEEE BMSB}, pp. 1-6, Jun. 2014.

\bibitem{SM}
R. Mesleh, H. Haas, S. Sinanovic, C. W. Ahn, and S. Yun, ``Spatial modulation,'' \emph{IEEE Trans. Veh. Technol.}, vol. 57, no. 4, pp. 2228-2241, Jul. 2008.

\bibitem{SM_mgz}
M. Di Renzo, H. Haas, A. Ghrayeb, S. Sugiura, and L. Hanzo, ``Spatial modulation for generalized MIMO: challenges, opportunities and implementation,'' \emph{Proceedings of the IEEE}, vol. 102, no. 1, pp. 56-103, Jan. 2014.

\bibitem{Massive_SM_ITA}
T. Narasimhan, P. Raviteja, and A. Chockalingam, ``Large-scale multiuser SM-MIMO versus massive MIMO,'' in \emph{Proc. ITA}, pp. 1-9, Feb. 2014.

\bibitem{Massive_SM_TVT}
P. Patcharamaneepakorn, S. Wu, C. Wang, H. M. Aggoune, M. M. Alwakeel, X. Ge, and M. Di Renzo, ``Spectral, energy and economic efficiency of 5G multi-cell massive MIMO systems with generalized spatial modulation,'' \emph{IEEE Trans. Veh. Technol.}, vol. 65, no. 12, pp. 9715-9731, Dec. 2016.

\bibitem{NOMA_SM_WXS}
X. Wang, J. Wang, L. He, and J. Song, ``Spectral efficiency analysis for downlink NOMA aided spatial modulation with finite alphabet inputs,'' \emph{IEEE Trans. Veh. Technol.}, vol. 66, no. 11, pp. 10562-10566, Aug. 2017.

\bibitem{Mmwave_SM_HLZ}
L. He, J. Wang, and J. Song, ``On generalized spatial modulation aided millimeter wave MIMO: spectral efficiency analysis and hybrid precoder design,'' \emph{IEEE Trans. Wireless Commun.}, vol. 16, no. 11, pp. 7658-7671, Nov. 2017.

\bibitem{Mmwave_SM_HLZ_TCOM}
L. He, J. Wang, and J. Song, ``Spatial modulation for more spatial multiplexing: RF-chain-limited generalized spatial modulation aided MmWave MIMO with hybrid precoding,'' \emph{IEEE Trans. Commun.}, to be publised.


\bibitem{SM_TBC}
B. Gong, L. Gui, Q. Qin, and X. Ren, ``Compressive sensing-based detector design for SM-OFDM massive MIMO high speed train systems,'' \emph{IEEE Trans. Broadcast.}, vol. 63, no. 4, pp. 714-726, Aug. 2017.

\bibitem{SM_Jiao}
T. Wang, S. Liu, F. Yang, J. Wang, J. Song, and Z. Han, ``Generalized spatial modulation-based multi-user and signal detection scheme for terrestrial return channel With NOMA,'' \emph{IEEE Trans. Broadcast.}, pp. 1-9, Oct. 2017.

\bibitem{Massive_SINR}
E. Bj\"{o}rnson, E. G. Larsson, and M. Debbah, ``Massive MIMO for maximal spectral efficiency: how many users and pilots should be allocated?'' \emph{IEEE Trans. Wireless Commun.}, vol. 15, no. 2, pp. 1293-1308, Feb. 2016.

\bibitem{yang2008information}
Y. Yang and B. Jiao, ``Information-guided channel-hopping for high data rate wireless communication,'' \textit{IEEE Commun. Lett.}, vol. 12, no. 4, pp. 225-227, Apr. 2008.

\bibitem{Wishart}
A. M. Tulino and S. Verd\'{u}, ``Random matrix theory and wireless communications,'' \emph{Foundations Trends Commun. Inf. Theory}, vol. 1, no. 1, pp. 1-182, Jun. 2004.

\end{thebibliography}
\end{document}